# On the Variability of the Solar Mean Magnetic Field: Contributions from Various Magnetic Features on the Surface of the Sun

Souvik Bose[1,2]  and K. Nagaraju[3]
[1] Institute of Theoretical Astrophysics, University of Oslo, P.O. Box 1029 Blindern, NO-0315 Oslo, Norway; souvik.bose@astro.uio.no
[2] Rosseland Centre for Solar Physics, University of Oslo, P.O. Box 1029 Blindern, NO-0315 Oslo, Norway
[3] Indian Institute of Astrophysics, Koramangala-2nd Block, Bangalore-560034, India

## Abstract

The solar mean magnetic field (SMMF) is referred to as the disk-averaged line-of-sight (LOS) magnetic field that also reflects the polarity imbalance of the magnetic field on the Sun. The origin of the SMMF has been debated over the past few decades, with one school of thought suggesting that the contribution to the SMMF is mostly due to the large-scale magnetic field structure, also called the background magnetic field, whereas other and more recent studies have indicated that active regions have a major contribution to the observed SMMF. In this paper, we re-investigate the issue of the origin of the SMMF by decomposing the solar disk into plages, networks, sunspots, and background regions, thereby calculating the variation in the observed SMMF due to each of these features. We have used full-disk images from *Solar Dynamics Observatory* (*SDO*)/AIA recorded at 1600 Å for earmarking plages, networks, and background regions and 4500 Å images for separating the sunspots. The LOS fields corresponding to each of these regions are estimated from the co-temporal *SDO*/Helioseismic and Magnetic Imager full-disk magnetograms. The temporal variation of the SMMF shows a near one-to-one correspondence with that of the background field regions, suggesting that they constitute the major component of the observed SMMF. A linear regression analysis based on the coefficient of determination shows that the background field dominates and accounts for 89% of the variation in the SMMF, whereas the magnetic field from the other features accounts for the rest 11%.

## 1. Introduction

Observing the Sun as a star provides information that is an average over the entire visible solar disk. The solar mean magnetic field (SMMF) is one such quantity representing the disk-averaged line-of-sight (LOS) magnetic field on the Sun. In the literature this is also known as the general magnetic field (Scherrer et al. 1977b) or the mean magnetic field (MMF; Plachinda et al. 2011) of the Sun. The SMMF essentially reflects the imbalance in the magnetic flux of opposite polarities on the visible disk (Svalgaard et al. 1975).

Observations of the SMMF first began at the Crimean Astrophysical Observatory (Severny et al. 1970). This was followed up most notably by the Wilcox Solar Observatory (Scherrer et al. 1977b) and the Mt. Wilson Observatory (Kotov et al. 1998). These measurements are generally made by letting unfocused light pass through a spectrograph slit (making sure that light from all parts of Sun is integrated) followed by measuring Zeeman splitting of the spectral lines under consideration using a Babcock-type magnetograph. Boberg et al. (2002), and more recently Kutsenko & Abramenko (2016), have shown that the SMMF can also be calculated by averaging the full-disk LOS magnetograms provided by the space-based Michelson Doppler Imager (Scherrer et al. 1995) on board the *Solar and Heliospheric Observatory* and the Helioseismic and Magnetic Imager (HMI; Schou et al. 2012) on board the *Solar Dynamics Observatory*, (*SDO*; Liu et al. 2012; Scherrer et al. 2012), respectively. This has significantly improved the chances of obtaining such measurements continuously more or less uninterrupted, which otherwise is limited by lack of continuous availability of clear skies in ground-based observations. However, the data obtained from both space- and ground-based observations complement each other for the study of short-term as well as long-term variation of the SMMF.

The availability of SMMF measurements for more than two Hale cycles has been very helpful in studying its temporal behavior on various timescales, ranging from a few days to several years. The SMMF exhibits a dominant periodicity of ≈27 days. The analysis by Haneychuk et al. (2003) reported that this rotation period does not vary with time, whereas a recent study by Xie et al. (2017) suggested that the rotation period of SMMF changes with the phase of the solar cycle. Besides, the amplitude of the SMMF varies slowly on the timescales of the solar cycle: being close to ±2 G during the sunspot maximum and ±0.2 G during the minimum of the solar activity (Plachinda et al. 2011). The SMMF variation, however, does not show direct relation with any of the solar cycle indices (Kotov 2008).

It is well established that the source of the interplanetary magnetic field (IMF) is located on the surface of the Sun (Wilcox & Ness 1965; Hudson et al. 2014). These source fields are organized into long-lived, large-scale structures known as solar sector structures. These sectors consist predominantly of unipolar fields, and the sector boundaries run almost parallel to the solar north–south direction. The sectors extend up to 35° latitude on both sides of the equator (Svalgaard & Wilcox 1975).

Based on the observed properties of the SMMF, it has been suggested that the contribution to and the variation in the imbalance of the opposite polarity fields is due to the large-scale structure, which is prominently bipolar in nature, and whose axis lies close to the equator (Severny 1971; Svalgaard & Wilcox 1975; Scherrer et al. 1977a; Hudson et al. 2014;

Xiang & Qu 2016; Gough 2017). However, a recent analysis by Kutsenko et al. (2017) suggests that the contribution of the sunspots toward the SMMF is more dominant whereas the role of the large-scale background field is negligible. They have divided the full-disk *SDO*/HMI magnetograms into three categories viz. strong (mainly comprising of sunspots), intermediate (plages and network fields), and the weak background fields and report that the strong and intermediate fields have major contribution to the SMMF. In this work, we follow the approach of decomposing the solar disk into sunspots, plages, and network features from intensity images through the conventional way of identifying them, meaning that the surface features are separated out based on the intensity (brightness) and area criterion. We make use of the *SDO*/AIA 1600 Å data to segregate various features like plages, enhanced and active network (AN), and *SDO*/AIA 4500 Å data for detection of sunspots on the solar disk. Subsequently, we mask these regions on the corresponding HMI LOS magnetograms to calculate the variation in the SMMF due to each of these features.

## 2. Data Description

We have used *SDO*/HMI magnetograms for the purpose of calculating the variation in the SMMF and full-disk images at 1600 and 4500 Å wavelengths recorded by *SDO*/AIA for detecting the corresponding solar surface features. HMI is one of the three instruments on board the *SDO* providing full-disk LOS magnetograms every 720 s (HMI.M_720s series) with a spatial resolution of $1''$ and a spatial sampling of $0''\!.5043 \times 0''\!.5043$ per pixel. AIA provides full-disk images of the Sun in extreme ultraviolet (94, 131, 171, 193, 211, 304, and 335 Å ) and ultraviolet (UV; 1600 and 1700 Å ) wavelengths with a spatial resolution of $1''\!.2$ and a plate scale of $0''\!.6$ per pixel and a temporal resolution of 12 s and 24 s, respectively. The 4500 Å visible light images have a cadence of nearly 1 hr. In the present study, we used one co-temporal LOS magnetogram from HMI.M_720s series, one full-disk image each from AIA 1600 and 4500 Å per day between 2011 March 23 and 2017 November 30.

The similarity of the observed chromospheric features in AIA 1600 Å with that of Ca II K motivated us to explore AIA 1600 Å full-disk images for the detection of plages and networks which otherwise have been identified and studied in images obtained mostly in Ca II H or K lines (Lefebvre et al. 2005; Bertello et al. 2010; Singh et al. 2012; Priyal et al. 2014; Chatterjee et al. 2016). In this paper, we quantitatively compare the observed features in AIA 1600 Å with the Ca II K data obtained from the Chromospheric Telescope (ChroTel). The ChroTel (Kentischer et al. 2008; Bethge et al. 2011) is a 10 cm robotic ground-based telescope located on Tenerife, Canary Islands. The ChroTel observes the Sun quasi-simultaneously in three channels, namely Ca II K, H$\alpha$ and He I 10830 Å . These channels are recorded with a gap of 10 seconds by using a Lyot filter on a 2k × 2k Kodak KAF-4320E CCD sensor. The Ca II K data is acquired in a single exposure mode with an exposure time of 1 s and filter pass band (FWHM) of 0.3 Å. The pixel scale corresponds to about $1''\!.02980 \times 1''\!.02980$.[4]

---

[4] The ChroTel data were downloaded from ftp://archive.leibniz-kis.de/pub/chrotel/lev1.0/.

## 3. Methods of Analysis

### 3.1. Automated Detection of Plages and Enhanced Network Regions from AIA 1600 Å images

Plage regions and enhanced network areas are surface features that are normally associated with active regions on the solar disk. In general, enhanced network regions are those that are created when the plages disperse into smaller regions (Priyal et al. 2014). They are found in close proximity to the plages, but are known to posses lesser magnetic flux.

The AIA data were first converted from level 1 to 1.5 by using the `aia_prep.pro` routine from the IDL SolarSoft package. We then accounted for the limb darkening (LD) effect by a median filtering process similar to Lefebvre et al. (2005), Bertello et al. (2010), and Chatterjee et al. (2016). We applied a 105 × 105 2D running median filter on each of the two images that were resized to 512 × 512. The size of the median filter matrix depends on its ability to smooth the reference image completely. Hence, it may vary from one data to the other since different data may have different intensity contrast. Further, resizing the original 4096 × 4096 data, quadrupled the computation speed.

Blurring the images helped in capturing the large-scale intensity variation across the disk. Subsequently, the filtered images were restored back to their original dimension before dividing the level 1.5 data with the derived LD profile. This helped in correcting the LD effect quite effectively. The radially symmetric polynomial fitting process for the LD profile as in Singh et al. (2012) or Priyal et al. (2014) was not used, primarily due to the instrumental issues. Moreover, it may not be able to correct for the varying dynamic range of intensity from one image to the other (Bertello et al. 2010).

The LD corrected images were subjected to intensity thresholding estimated by an adaptive thresholding technique (Niblack 1985) based on their mean ($\mu_{\text{image}}$) and the standard deviation ($\sigma_{\text{image}}$) given by $I_{\text{crit}} = \mu_{\text{image}} + K\sigma_{\text{image}}$. The purpose was to generate binary images whose pixels have intensity values greater than $I_{\text{crit}}$. The intensity threshold varies with the instrument used at different observatories and it also depends on the FWHM of the filtergraph used (Priyal et al. 2014). We have chosen the value of $K$ to be 1.71 for the AIA data after repeated experimentation with several different values on a large number of data sets.

Subsequently, an area thresholding criterion was also imposed on each of the binary images so as to distinguish and clearly detect the plages and the enhanced network regions on the disk. We followed a simple region labeling technique where the pixels whose intensity values were greater than the $I_{\text{crit}}$ (as described in the previous paragraph) and occupied an area greater than 0.5 arcmin$^2$ were classified as plages whereas those that have intensity values greater than $I_{\text{crit}}$ but occupied an area between 0.1 arcmin$^2$ and 0.5 arcmin$^2$ were classified as enhanced networks. A similar area criterion was also used by Priyal et al. (2014). A closing morphological operation with a 3 × 3 kernel was also applied to eliminate small dark background noise from the foreground image. The result of the detection algorithm is shown in panel (A) of Figure 1 with the contours of detected plages and enhanced networks overplotted on an image obtained by *SDO*/AIA on 21.04.2012 at 00:00:00 at 1600 Å. The red and the blue contours denote the plages and enhanced networks, respectively.

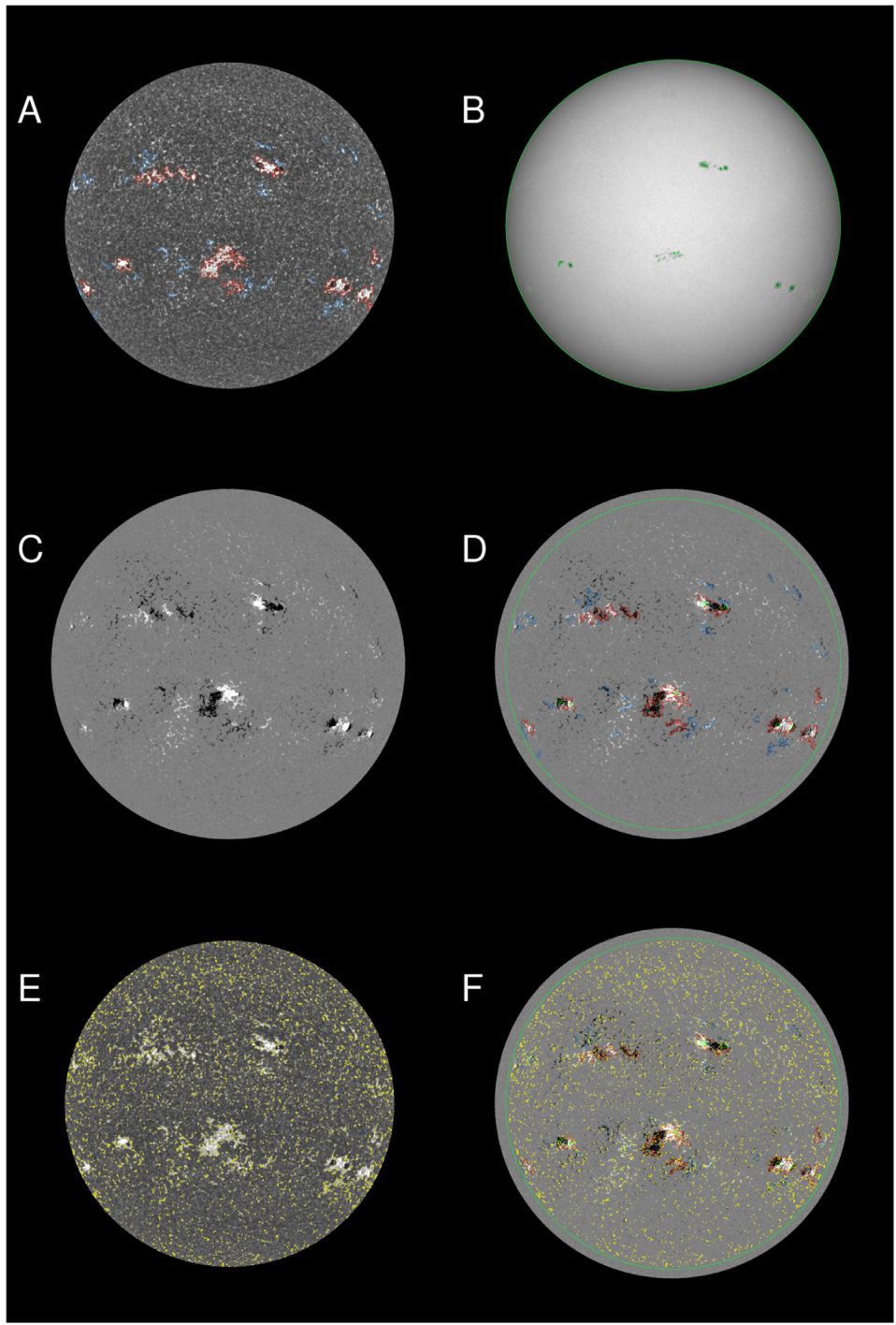

**Figure 1.** Sample of full-disk images of the Sun observed on 21.04.2012 00:00:00 UTC by *SDO*/AIA at 1600 and 4500 Å and a corresponding *SDO*/HMI magnetogram with the contours of the identified features overlaid on them. Different panels in this figure correspond to (A): AIA 1600 Å image with contours of Plages (red) and Enhanced Networks (blue); (B): AIA 4500 Å image with contours of sunspots (fluorescent green); (C): HMI magnetogram co-aligned with AIA; (D): contour plots of plages, enhanced networks, and sunspots on the magnetogram; (E): contour plots of active networks on AIA 1600 Å image (yellow); and (F): all the contours overplotted on the magnetogram (scaled between −100 G and 100 G). The gray regions corresponds to what we call the background regime. The circular fluorescent green contour in panels (D) and (F) depicts the solar disk within 0.97 $R_\odot$.

### 3.2. Automated Detection of AN Regions from AIA 1600 Å and Sunspots from AIA 4500 Å Images

In this section we discuss the automated detection of sunspots and ANs from AIA 4500 and 1600 Å wavelengths, respectively. Sunspots are the regions of strong magnetic fields and they appear darker than the surrounding regions. They are also very dynamic in nature.

We have used AIA 4500 Å data because sunspots have a good contrast when viewed in white light, thereby facilitating better detection. Also, both the 1600 and 4500 Å data sets are recorded from the same instrument and thereby have the same pixel scale. We also made use of the co-temporal data sets to prevent any misidentification of features on the solar disk. For the detection, we considered LD corrected AIA 4500 Å images by smoothing it with a 40 × 40 2D median filter, and proceeded in a similar manner described in the previous section. We then simply employed the adaptive intensity thresholding technique given by $I_{\text{crit}}$, by considering pixels lesser than $\mu_{\text{image}} + 0.4\sigma_{\text{image}}$ as sunspot pixels. The value of $K$ was chosen after rigorous experimentation and trial on a number of images at different periods of the solar cycle. Figure 1(B) demonstrates the automated detection of sunspots.

AN regions are thought to be created either by emerging magnetic fields or by further decay of plages (Foukal et al. 2009; Priyal et al. 2014). The fluxes are relatively weaker, but they can be located anywhere on the solar disk, unlike active regions (Priyal et al. 2014). Many, like Zwaan (1987) and Worden et al. (1998), have used Ca II K data for the detection of AN with a slightly lower intensity threshold than for plage and enhanced network regions. We also used a similar intensity criteria to identify them but, from the AIA 1600 Å data. We again considered the same data those were used for the detection of plages and enhanced network as described in the previous section, and applied the adaptive intensity thresholding technique after accounting for the LD correction. We found that pixels with intensities in the range between $\mu_{\text{image}} + 1.65\sigma_{\text{image}}$ and $\mu_{\text{image}} + 1.71\sigma_{\text{image}}$ resembled that of the ANs in the AIA 1600 Å data. This is shown in Figure 1(E) in yellow contours. Our detection of ANs is comparable to that of Figure 8(D) in Priyal et al. (2014).

It is worthwhile to note here that the adaptive intensity thresholding technique that we employed in our detection is much more robust compared with constant global thresholding technique utilized in Singh et al. (2012) and Priyal et al. (2014). This is because our threshold value adapts itself to the basic properties of the images, which are given by $\mu$ and $\sigma$. This automatically takes into account the variation in intensity contrast from one image to the other. Different values of the area thresholding criterion ranging from 0.25 arcmin$^2$ to 1 arcmin$^2$ have already been inculcated in plage detection algorithms in the recent past (Priyal et al. 2014; Chatterjee et al. 2016). We find that, with the chosen value of 0.5 arcmin$^2$, the feature detection algorithm worked satisfactorily in our case and the numerical difference between the area thresholds may be attributed to the different instruments used for recording the data.

Once these regions have been identified and separated out, we overplot the contours (masking) obtained by the detection techniques described above on the corresponding HMI.M_720s series LOS magnetograms. However, before this, the original data obtained from HMI needed to be aligned and re-scaled with the AIA data sets for proper comparison. This was done quite effectively by reading the HMI.M_720s data with `read_sdo.pro` and further processing them with `aia_prep.pro` routines from the Solar Software package of IDL.[5] Figure 1(C) shows the magnetogram co-aligned with the AIA data. Figure 1(D) shows the re-scaled LOS magnetogram with the contours of plages, enhanced networks and the sunspots in red, blue and florescent green, respectively. Finally, in Figure 1(F), we demonstrate the complete detection of all the surface features on the magnetogram. We clearly see a high degree of accuracy in detecting the surface features which closely traces the magnetic features on the Sun.

After identifying the surface features, the corresponding binary masks were generated for every single image between 2011 March 23 and 2017 November 30 at a temporal cadence of 1 image per day recorded co-temporally at UT ≈ 00:00:00 hr. We then grouped the masks into three main categories: (1) sunspots, (2) plages, enhanced and ANs (as one entity), and (3) background regions that do not belong to either (1) or (2). To calculate the percentage variation in the SMMF due to each of these regions, we used the coefficient of determination ($R^2$; $R$ = Pearson's correlation coefficient) method based on a linear regression analysis. We began by calculating the total signed magnetic field of the pixels corresponding to the sunspot, plage-network and background regions, separately. The total field from each of the identified regions was divided by the total number of pixels ($N^T$) corresponding to the full solar disk. Naturally, the sum of the normalized total fields of all the three regions will equate to the SMMF, which is illustrated in the following equation:

$$\text{SMMF} = \frac{B_{\text{SS}}^T + B_{\text{PN}}^T + B_{\text{BG}}^T}{N^T}, \quad (1)$$

where $B_{\text{SS}}^T$, $B_{\text{PN}}^T$ and $B_{\text{BG}}^T$ represent the signed total field of sunspots, plages-networks, and background regions, respectively. The normalized total field corresponding to the individual category of pixels essentially represents the mean field of the corresponding region weighted by the ratio of the number of pixels covered by each of these regions to the total number of pixels of the full disk. In other words Equation (1) can be rewritten as

$$\text{SMMF} = \frac{N_{\text{SS}} * B_{\text{SS}}^M}{N^T} + \frac{N_{\text{PN}} * B_{\text{PN}}^M}{N^T} + \frac{N_{\text{BG}} * B_{\text{BG}}^M}{N^T}, \quad (2)$$

where, $B_{\text{SS}}^M$, $B_{\text{PN}}^M$, and $B_{\text{BG}}^M$ represent the MMF components, and $N_{\text{SS}}$, $N_{\text{PN}}$, and $N_{\text{BG}}$ represent the number of pixels occupied by the sunspots, plages-networks and background regions, respectively. For simplicity we write Equation (2) as

$$\text{SMMF} = \bar{B}_{\text{SS}} + \bar{B}_{\text{PN}} + \bar{B}_{\text{BG}}, \quad (3)$$

where $\bar{B}_{\text{SS}}$, $\bar{B}_{\text{PN}}$ and $\bar{B}_{\text{BG}}$ represent weighted-mean field of the sunspots, plages-networks, and background regions, respectively. A scatterplot analysis based on a linear regression technique was then employed to quantify the variability in the SMMF due to $\bar{B}_{\text{SS}}$, $\bar{B}_{\text{PN}}$, and $\bar{B}_{\text{BG}}$, with the help of the coefficient of determination method.

### 3.3. Comparison of the Feature Detection between AIA 1600 Å with Ca II K

For the sake of completeness, we now compare the detection of chromospheric features like plages, enhanced and ANs from AIA 1600 Å with Ca II K data obtained co-temporally with the ChroTel on 2015 August 3 T-16:00:00 UTC. The Ca II K data was co-aligned with the corresponding AIA level 1.5 1600 Å image by (1) correcting for the solar north for both the images to ensure the angle of rotation was perfectly aligned with respect to each other; (2) re-sampling the ChroTel Ca II K images to match the spatial scales of the AIA images of 0″.6 per pixel; (3) computing the shifts between the two by using standard cross-correlation algorithm and compensating them; and (4) finally by degrading the spatial resolution of AIA 1600 Å to bring it down to the ChroTel resolution so as to have an efficient comparison between the two detections. This co-alignment and reduction process proved to be accurate down to the level of the AIA pixel scale and the final size of both the images was 4096 × 4096.

---

[5] As described in https://www.lmsal.com/sdodocs/doc/dcur/SDOD0060.zip/zip/entry/.

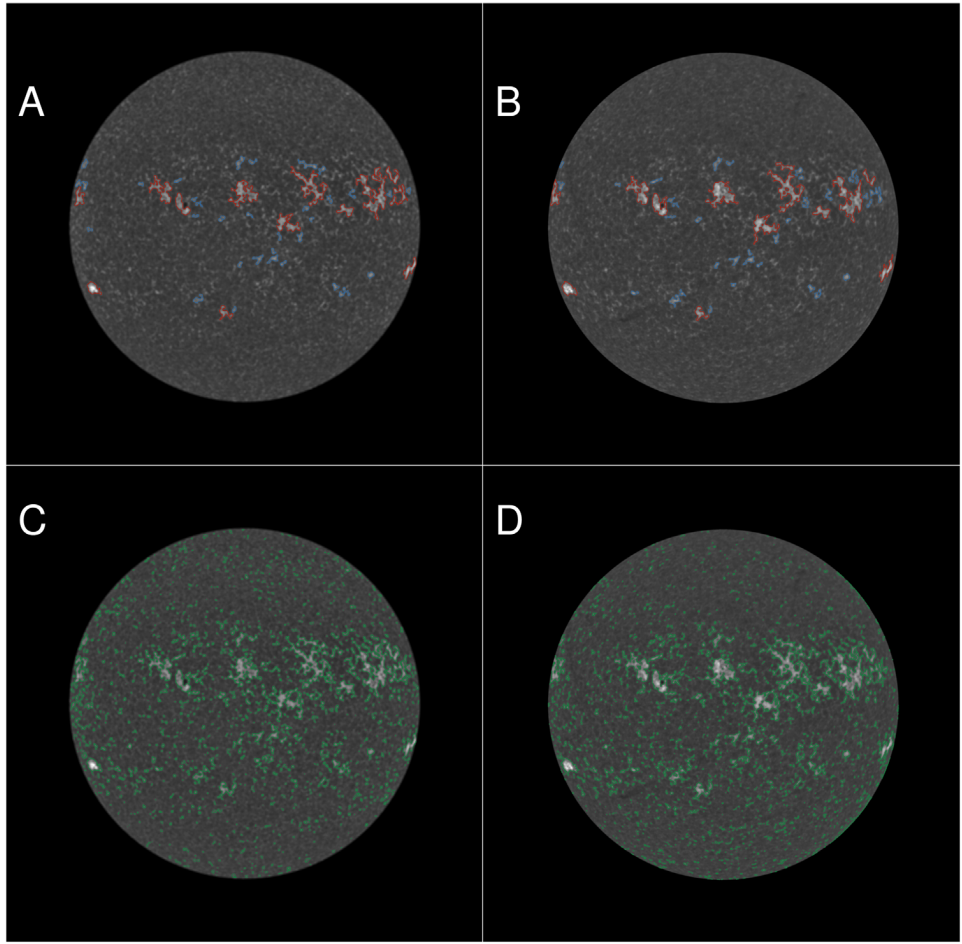

**Figure 2.** Co-temporal full-disk images of the Sun at 1600 Å (brought down to ChroTel resolution), (panels (A) and (C)), and 3934 Å Ca II K (panels (B) and (D)) as observed on 03.08.2015 T-16:00:00 UTC by *SDO*/AIA and ChroTel, respectively. Panels (A) and (B) compare the detection of the plages (red contours) and the enhanced networks (blue contours), and panels (C) and (D) compare the active network detection between them (green contours).

Subsequently, we corrected for the LD profile of both the images and followed the adaptive intensity thresholding criterion with values of $K$ equal to 1.71 and 1.72 for AIA and ChroTel data respectively. Further, an area threshold with exactly the same value of 0.5 arcmin$^2$ was imposed for the detection of plages, and pixels occupying an area between 0.1 arcmin$^2$ and 0.5 arcmin$^2$ were considered as enhanced networks. Figure 2(A) shows the plages (red contours) and the enhanced networks (blue contours) detected with AIA 1600 Å data compared with Ca II K data in Figure 2(B). The AN (in green contours) is also compared in Figures 2(C) and (D) for AIA 1600 Å and ChroTel Ca II K data, respectively. Pixels with intensities between $\mu_{\mathrm{image}} + 1.43\sigma_{\mathrm{image}}$ and $\mu_{\mathrm{image}} + 1.71\sigma_{\mathrm{image}}$ in AIA 1600 Å and between $\mu_{\mathrm{image}} + 1.41\sigma_{\mathrm{image}}$ and $\mu_{\mathrm{image}} + 1.72\sigma_{\mathrm{image}}$ in ChroTel 3934 Å were identified as ANs on the solar disk in this case.

Computation of the combined area occupied by the plages, enhanced and ANs with both the data sets yielded very similar results. The sum of the areas equated to 246707.30 arcsec$^2$ in AIA 1600 Å and 250528.34 arcsec$^2$ in the Ca II K ChroTel image and they differed by less than 1.5%. These numbers are also found to be well in accordance with the results obtained by Priyal et al. (2014) for the previous solar cycles that further validated our detection algorithm. The observed similarities between the two data sets, and the fact that the AIA 1600 Å data can be obtained co-temporally with the HMI magnetograms without seeing hindrances, encouraged us to use the space-based AIA 1600 Å images for the long-term identification of the different features.

### 4. Results and Discussion

Plots of the SMMF and the weighted MMFs corresponding to the different features (as in Equation (3)) on the solar surface are shown in Figure 3 as a function of time, for the entire analysis period. Figure 3(A) shows the peak value of the SMMF to be about 2.5 G that occurred in 2014 December and is consistent with Kutsenko & Abramenko (2016). The dotted vertical line corresponds to 2014 April, which corresponds to the maximum of solar cycle 24. The background field component in Figure 3(B) emulates the SMMF quite distinctly, notably also peaking co-temporally with the SMMF. There exists a clear visual correlation between the two. On the other hand, variation in the LOS field corresponding to other regions such as plages, networks, and sunspots appears to be completely uncorrelated with the SMMF (Figures 3(C) and (D)). This is in agreement with the earlier works of Severny et al. (1970), Svalgaard (1972), Kotov et al. (1977), and Scherrer et al. (1977a), which supports the view that the large-scale background magnetic field structures on the Sun supply the leading contribution toward the variation in the SMMF. In our investigation, they resemble the unbalanced background field regimes.

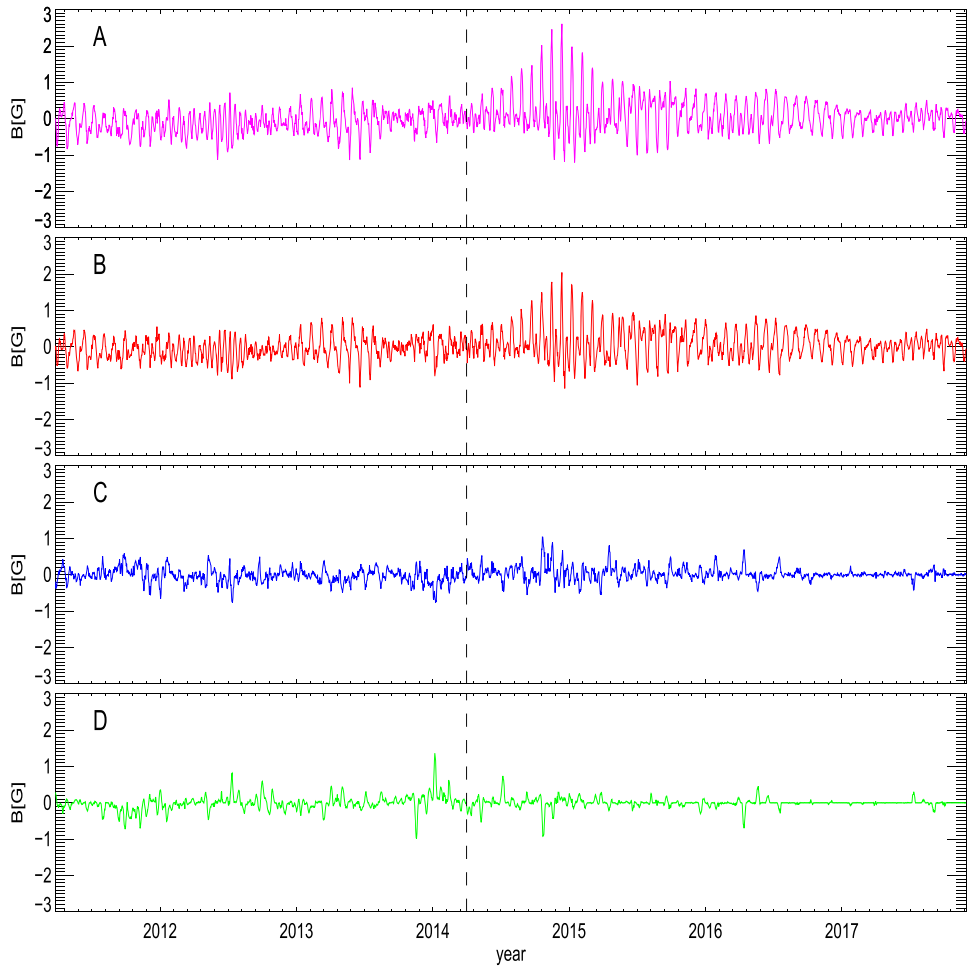

**Figure 3.** Plots of SMMF (panel (A)); the weighted-mean fields of background (panel (B)); plages, enhanced networks, and active networks (panel (C)); and sunspots (panel (D)). The dotted vertical line corresponds to the time of solar maximum of cycle 24, which was during 2014 April.

It is also important to mention that the peak of the SMMF and the background field component appears later than the peak of the solar cycle. Sheeley & Wang (2015) have investigated this in their paper where they report that there is a systematic shift in the peak of the SMMF with respect to the solar cycle (peaks) also in the previous cycles.

We follow the coefficient of determination method as discussed in Section 3 to calculate the percentage variation in the SMMF with linear regression fits on the scatter plots as shown in Figure 4. The coefficient of determination, as given by $R^2$, provides an estimation of the variation in the dependent variable that is predicted by the independent variable. In other words, $R^2 \times 100$ variation in the y-variable (dependent) is predicted by the x-variable (independent). The left and the right panels of Figure 4 shows the relationship between the SMMF and the weighted-mean background field ($\bar{B}_{BG}$) and the weighted-mean plage-network fields ($\bar{B}_{PN}$), respectively. We find that only 9.9% variation in the SMMF is due to $\bar{B}_{PN}$ whereas about 88.9% variation is due to the $\bar{B}_{BG}$. Performing a student's t-test on $R$, revealed that the correlation between the $\bar{B}_{SS}$ and the SMMF is statistically insignificant at 95% confidence level and hence we have not reported it here. Therefore, it is clear that most of the variation in the SMMF is mainly due to the large-scale background fields on the solar disk and the active regions, in particular the sunspots, have little contribution toward the same.

The above finding is in contrast with Kutsenko et al. (2017) who reported that the strong-flux component, comprising of the active regions, often exhibits amplitude variations similar to that of the SMMF. In other words, we find that the contribution of the active regions toward the variation in the SMMF is relatively weak compared to the large background areas (Figure 4). For further investigation, we also performed the decomposition of the features (see the Appendix) corresponding to the HMI magnetogram on 2014 December 13-05:24 UTC (this corresponded with the date chosen by Kutsenko et al. 2017) with our automated detection code. We quite successfully identified the features that are very similar to what they referred to as the active region magnetic flux concentrations in their paper (see Figure 2 in Kutsenko et al. (2017) and Figure 5 in this paper). Despite this, we find that the background region plays a quintessential role in the SMMF variability. There could be two possible explanations for this difference. (1) Due to the LOS nature of the HMI magnetograms, solar features near the disk center will have different magnetic flux values compared to limb. Decomposition on the basis of magnetic flux might therefore lead to false identification of features. (2) Thresholding the magnetograms with the magnetic fluxes may detect some regions or pixels that are strong, but are not physically part of any active region. Those may mistakenly be identified as strong-flux regions. Our detection method on the other hand, does not suffer from the above limitations. LD correction of the intensity images helps in proper detection of features even in the limb or the edges of the disk and the intensity followed by area thresholding criterion restricts misidentification of features to a large extent.

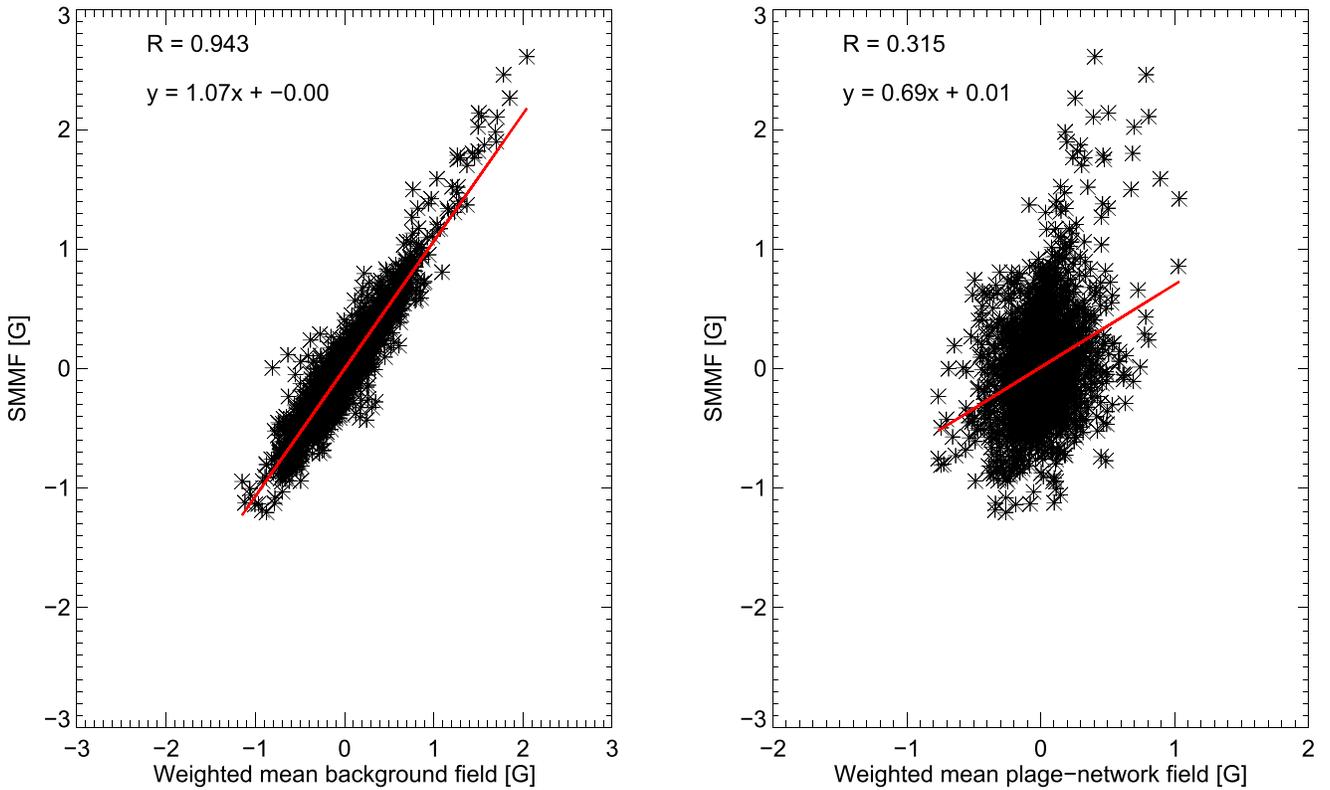

**Figure 4.** Scatterplot representation to estimate the variability of the background and the plage (and network) regimes. Left panel: comparison of the SMMF with the weighted-mean background field. Right panel: similar comparison between the plage and network field with the SMMF. $R$ is the Pearson's coefficient and the red line indicates the best-fit straight line through the data.

## 5. Conclusion

The uninterrupted seeing-free data from *SDO* has been analyzed in order to obtain the contribution from various magnetic features on the surface of the Sun to the observed variability in the SMMF. Surface features such as plages, networks, and sunspots were identified using *SDO*/AIA full-disk intensity images at 1600 and 4500 Å and the corresponding weighted-mean LOS field was calculated from the co-temporal *SDO*/HMI LOS magnetograms.

A comparison of the features detected between Ca II K and AIA 1600 Å images revealed a near one-to-one correspondence for the same intensity and area criteria, suggesting that the latter images can be effectively used for long-term detection of plage regions and network areas exploiting the continuous and homogeneous AIA data archive.

The variation in the SMMF due to the weighted-mean fields corresponding to the different features, including the background, were computed. A close look at the temporal variation, along with $R^2$, and the comparison with the mean field of other regions suggests that the background field is the major contributor to the variability in the SMMF. Active regions, including the AN regimes, show no or very little correlation whatsoever. In particular, the contribution from the sunspots is random and statistically insignificant. This is established with the help of the coefficient of determination method with linear regression fits on the scatter plots. We found that the background field component contributed about 88.9% whereas the contribution of the plages and the network field was about 9.9% toward the SMMF variability. Based on these findings, we conclude that the origin of the observed variability in the SMMF lies in the polarity imbalance of a large-scale magnetic field structures on the visible surface of the Sun.

Further, we would like to remark that the presence of sunspot activity on the surface of Sun may influence the amplitude of the SMMF, that is observed to change from about 0.5 G during solar cycle minimum to ≈2 G around solar cycle maximum. However, sunspots do not directly contribute to the observed SMMF, as is clearly shown in this work.

We would like to thank the anonymous referee for the critical comments that helped in fine-tuning the presentation of the results. We also acknowledge Luc Rouppe van der Voort for his careful reading of the manuscript that enhanced its readability. *SDO* is a mission for NASA's Living With a Star (LWS) program. The *SDO*/HMI data were provided by the Joint Science Operation Center (JSOC). S.B. would like to acknowledge the support by the Research Council of Norway,

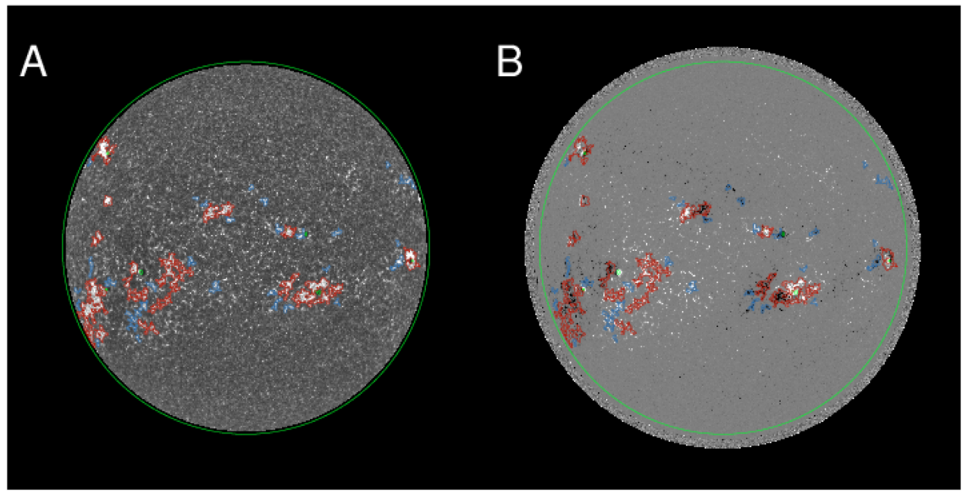

**Figure 5.** Features detection from the data set recorded on 13.12.2014. (A): AIA 1600 Å image with the plages (red), enhanced network (blue), and sunspot (fluorescent green); (B): corresponding HMI magnetogram with all the contours overlaid on it. The fluorescent green circle corresponds to 0.97 of solar radius.

## Appendix
## Detection of Features Corresponding to the HMI Magnetogram on 13.12.2014-05:24 UTC

In this appendix, we highlight our detection criterion for the magnetogram and the AIA data set corresponding to Figure 2 in Kutsenko et al. (2017). We used the same criterion for the detection of plages, enhanced networks, and sunspots as described in the text, and we overlay them on the corresponding HMI.M_720s magnetogram. It is clear from the Figure 5 that our detection of plages, enhanced networks, and sunspots (combined) give a similar impression with that of the active region flux concentrations in their paper.

### ORCID iDs

Souvik Bose  https://orcid.org/0000-0002-2180-1013